\documentclass[11pt,twoside]{article}
\usepackage{asp2004}
\usepackage{psfig}
\usepackage{epsf}
\usepackage{graphics}
\usepackage{lscape}
\newcommand{\z}{$z$}
\newcommand{\percc}{cm$^{-3}$}
\newcommand{\column}{cm$^{-2}$}
\newcommand{\halpha}{H$\alpha$}
\newcommand{\HI}{H$\,${\small I}}
\newcommand{\HII}{H$\,${\small II}}

\begin{document}
\title{Extraplanar Dust: a Tracer of Cold Dense Gas in the Thick Disks of Spiral Galaxies}
\author{J. Christopher Howk}
\affil{Ctr. for Astrophysics \& Space Sciences, 
  Univ. of California, San Diego, 
  MS-0424, La Jolla, CA 92014, USA}

\begin{abstract}
  
  The interstellar thick disks of galaxies contain not only gas, but
  significant quantities of dust.  Most of our knowledge of
  extraplanar dust in disk galaxies comes from direct broadband
  optical imaging of these systems, wherein the dust is identified due
  to the irregular extinction it produces against the thick disk and
  bulge stars.  This observational technique is sensitive to only the
  most dense material, and we argue much of the material identified in
  this way traces a cold phase of the interstellar thick disks in
  galaxies.  The presence of a cold, dense phase likely implies the
  interstellar pressures in the thick disks of spiral galaxies can be
  quite high.  This dense phase of the interstellar medium may also
  fueling thick disk star formation, and \halpha\ observations are now
  revealing \HII\ regions around newly-formed OB stars associations in
  several galaxies.  We argue that the large quantities of dust and
  the morphologies of the structures traced by the dust imply that
  much of the extraplanar material in disk galaxies must have been
  expelled from the underlying thin disk.

\end{abstract}
\thispagestyle{plain}

\section{Introduction}

Understanding the structure of the gaseous component of disk galaxies
is an important step for unraveling their on-going evolution. In
particular, extraplanar gas in galaxies is a potentially-important
probe of the effects of kinetic and radiative feedback from massive
stars to the gas in galaxies, as well as of the role that the
accretion of metal-poor gas plays in the evolution of modern galaxies.
Most studies of extraplanar matter have focused on the gas content of
the interstellar thick disks and halos of galaxies.  However, the
processes that transport matter from the thin disks of galaxies into
the thick disks and extended halos will act on both gas and
interstellar dust grains.  The presence of extraplanar dust in
galaxies can strongly affect the thermal balance of the gas, and it
will definitely affect an observer's view of a galaxy through its
impact on the transfer of radiation.

The presence of extraplanar dust in the canonical edge-on galaxy NGC
891 was noted many years ago (Sandage 1961; Dettmar 1990; Keppel et
al.  1991).  However, an analysis of the implications of the extensive
web of dusty extraplanar material in this galaxy did not occur until
recently (Howk \& Savage 1997, 2000).  In the last few years,
extraplanar dust has been recognized and studied in many edge-on
systems (Sofue, Wakamatsu, \& Malin 1994; Howk \& Savage 1999; Alton
et al. 2000; Rossa \& Dettmar 2003; Thompson, Howk, \& Savage 2004),
and we now know that the presence of extraplanar dust and gas are
indeed coupled.

In this contribution, I will concentrate on the physical
interpretation of the observed extraplanar dust and its implications
for the nature of the interstellar thick disks of spiral galaxies.  In
particular, I will summarize some of the evidence that material
identified with extraplanar dust may represent a dense, cold medium in
the thick disks of galaxies.

\section{Extraplanar Dust in Spiral Galaxies}

\begin{figure}[!ht]
%\epscale{0.8}
\plotone{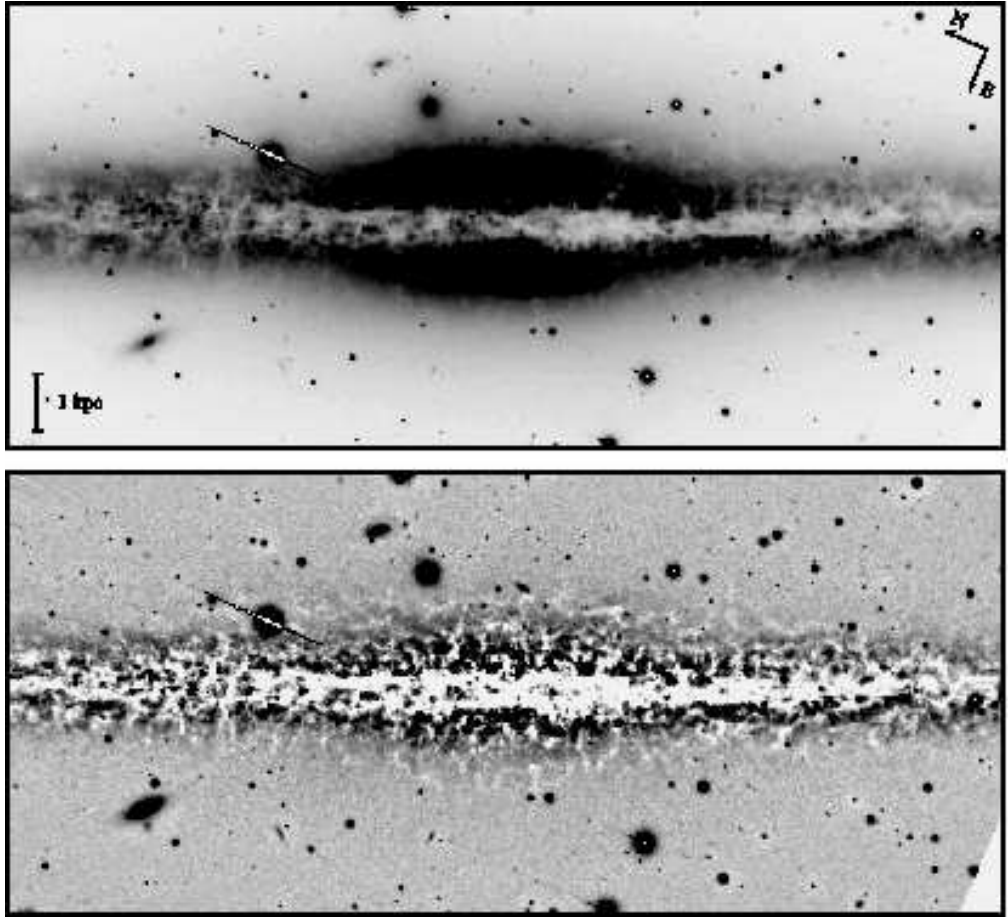}
\caption{Two versions of a broad-band optical (V-band) image of the edge-on
  galaxy NGC 891 from Howk \& Savage (2000).  The top panel shows the
  direct V-band image.  The bottom panel shows an unsharp-masked
  version of the V-band image.  The latter is produced by smoothing
  the original image and dividing the original by this smoothed
  version.  The purpose is to remove large-scale surface brightness
  variations, e.g., due to the vertically-decaying light from the
  stellar disk or from the bulge of the galaxy.  
\label{fig:n891}}
\end{figure}

It is now established that at least $\sim40\%$ of spiral galaxies in
the local Universe show some form of extraplanar (thick disk) dust
and, hence, gas (Rossa \& Dettmar 2003; Howk \& Savage 1999).  The
most common technique currently employed for identifying extraplanar
dust is to image directly edge-on spiral galaxies in the optical,
searching for obvious evidence of shadowing of the thick disk, halo,
and bulge stars by optically-thick foreground dust.  Figure
\ref{fig:n891} gives an example of this approach, showing two views of
the V-band image of NGC 891 from Howk \& Savage (2000).  The presence
of dust in the thick disk of this galaxy is obvious from these images.

%%%%%%%%%%%%%%%%%%%%%%%%%%%%%%%%%%%%%%%%%%%%%%%%%%%%%%%%%%%%%%%%%%%%%%

%%%%%%%%%%%%%%%%%%%%%%%%%%%%%%%%%%%%%%%%%%%%%%%%%%%%%%%%%%%%%%%%%%%%%%

From the stand-point of total exposure time required, such broadband
imaging is by far the most efficient manner of finding extraplanar
material in galaxies (although it does require high-resolution --
$\la1\arcsec$ -- imaging).  The presence of extraplanar dust is likely
a flag that a significant amount of extraplanar gas is present, as
well.  Not only is there likely to be a significant amount of gas
associated with the dust structures detected through direct optical
imaging, but the surveys of Rossa \& Dettmar (2003) and Howk \& Savage
(1999) have shown that $\sim90\%$ of galaxies exhibiting extraplanar
dust also have extraplanar diffuse ionized gas (DIG) detectable
through \halpha\ emission.

While it is straightforward to note the presence of extraplanar dust
from images such as that shown in Figure \ref{fig:n891}, there are
several biases inherent to this approach.  As discussed in Howk \&
Savage (2000) and more recently in Thompson et al. (2004), we are only
able to detect dusty regions through direct optical imaging because of
they have significantly lower surface brightness than their
surroundings.  This implies that the dust-bearing clouds seen in our
images must have a higher column density of dust (and gas) than their
surroundings: a smooth distribution of dust would produce no contrast
and would be undetectable in these images.  In practice this likely
implies the dusty clouds are more dense than their surroundings
(assuming the thick disk gas has a relatively uniform dust content).
A number of other, less scientifically interesting, biases should be
considered when looking at images such as those in Figure
\ref{fig:n891}.  The requirement of large contrast tends to bias us
toward detecting dust on the near side of galaxies, and, due to
signal-to-noise constraints, we are more likely to detect high-\z\ 
dust in regions of intrinsically higher surface brightness (e.g.,
against the bright light of a galactic bulge or at lower heights above
the midplane).

Using estimates of the ``apparent extinctions'' produced by individual
thick disk dust clouds, Howk \& Savage (1997, 1999, 2000) and Thompson
et al. (2004) have estimated physical properties of these clouds,
albeit crudely.  Because the apparent extinctions will always
underestimate the true extinctions through the clouds (see Howk \&
Savage 1997), all of the physical quantities derived using the
apparent extinctions are lower limits.  Assuming the dust-to-gas ratio
in these clouds is similar to that found in the disk of the Milky Way
(which is probably not too bad an assumption; Thompson et al. 2004),
the dusty cloud complexes detected in our images must have
$N(\mbox{\HI}) \ga 10^{20}$ \column.  Furthermore, the densities in
these clouds must be quite high.  Examining the apparent extinctions
and sizes of the smallest structures in the extraplanar cloud
complexes in NGC 891 as seen in new images from the Advanced Camera
for Surveys on board the {\em Hubble Space Telescope} suggests the
densities in these clouds may be $n_{H} \ga 25$ \percc.

The combination of these estimated column densities and the projected
sizes of the cloud complexes in our images imply total masses of
$\sim10^4$ to $10^5$ M$_\odot$ or higher {\em in each complex}.  Such
masses are consistent with those of the individual giant molecular
clouds in the Milky Way.  In galaxies with detectable extraplanar
dust, we typically find hundreds of absorbing structures at $z\la2$
kpc all along the central regions of the disk (within $R\la8$ kpc,
similar to the radial extent of detectable DIG).  Howk \& Savage
(1997) estimated the total mass of the ensemble of clouds in NGC 891
to be $\sim10^8$ M$_\odot$, comparable to the total mass of
extraplanar DIG material (Dettmar 1990).

\begin{figure}[!ht]
%\epscale{0.8}
\plotone{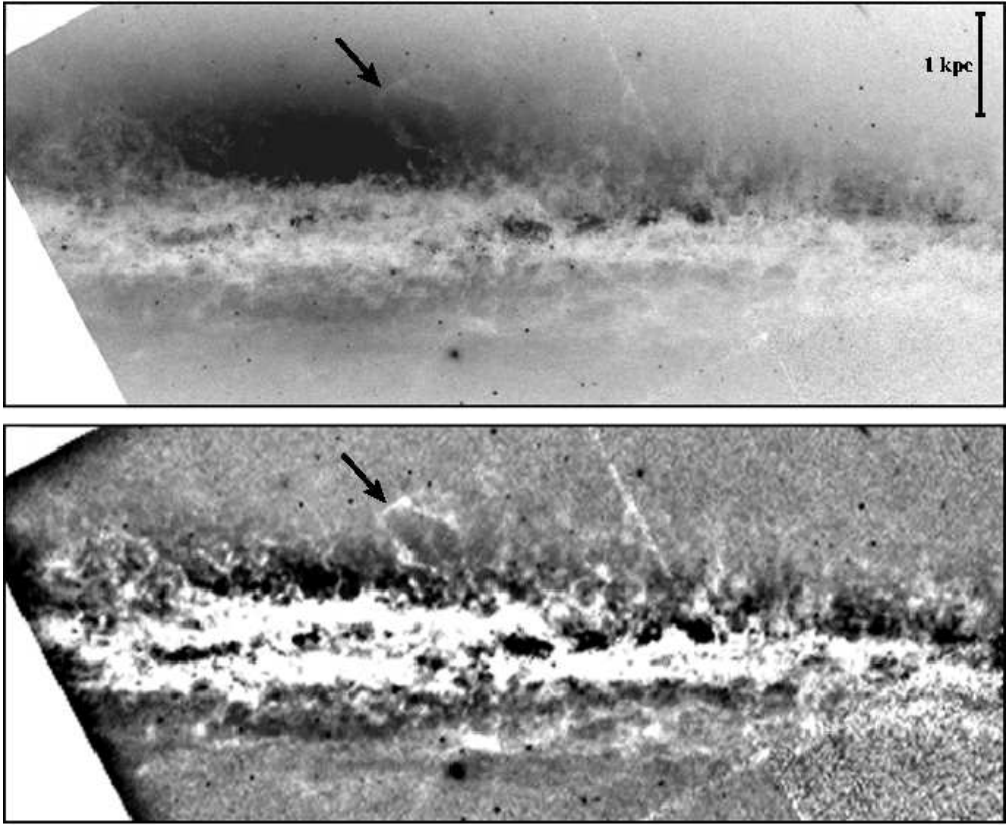}
\caption{Two versions of a broad-band optical (B-band) image of the edge-on
  galaxy NGC 4217 from Thompson et al. (2004) taken with the Wide
  Field and Planetary Camera 2 on board {\em HST}.  The top panel
  shows the direct B-band image.  The bottom panel shows an
  unsharp-masked version of the B-band image.  The large loop
  discussed in the text is marked with an arrow.
\label{fig:n4217}}
\end{figure}

Morphologically, the extraplanar dust structures seen in direct
optical images are quite complex and varied.  Much of the complexity
is likely due to the observed clouds residing at different depths
through a galaxy.  The identification of cloud complexes which may
represent coherent structures is extremely difficult at heights
$z\la1$ kpc from the midplanes of spirals, particularly toward the
centers of galaxies where more structures may be present along a given
sight line.  In fact, we believe that sight lines through an edge-on
galaxy with extraplanar dust are typically optically thick for heights
$z\la1$ kpc.  At such heights, every sight line intercepts at least
one dust-bearing cloud, each of which we believe to have $A_V > 1$.
At larger heights, where the confusion is significantly lessened, it
is possible to identify what appear to be individual, sometimes
isolated clouds or structures.  As an example of this, Figure
\ref{fig:n4217} shows a portion of our {\em HST} image of NGC 4217
(Thompson et al.  2004).  Thompson et al. note the presence of a
structure that appears to be a large loop (marked in Figure
\ref{fig:n4217}) with diameter $\sim800$ pc centered at $z\sim1300$ pc
from the midplane of this galaxy.  Such a structure would be confused
at lower heights from this or other galaxies.  Even at the height of
this structure, though, one worries that what appears to be a loop may
be caused by several overlapping, but unrelated, cloud complexes.

%%%%%%%%%%%%%%%%%%%%%%%%%%%%%%%%%%%%%%%%%%%%%%%%%%%%%%%%%%%%%%%%%%%%%%

%%%%%%%%%%%%%%%%%%%%%%%%%%%%%%%%%%%%%%%%%%%%%%%%%%%%%%%%%%%%%%%%%%%%%%

\section{Cold, Dense Gas in the Multiphase Thick Disks of Spiral Galaxies}

We have argued that the dust-laden clouds identified in the thick
disks of spiral galaxies through direct optical imaging represent a
cold, dense phase of the thick disk ISM (Howk \& Savage 1999, 2000).
The derived column densities and particle densities are consistent
with this picture.  The high column densities we derive for many
individual structures are high enough that they would have molecular
fractions of $>25\%$ if the physical conditions (i.e., dust content
and radiation fields) are similar to those found in the disk of the
Milky Way.  There are other observational indications that, sometimes
indirectly, lead us to believe the thick disks of galaxies contain a
CNM that is traceable through its extinction.

First, it is clear that the interstellar thick disks of spiral
galaxies are multiphase media.  The observations of gas at
temperatures that span several orders of magnitude (e.g., through
observations of \halpha\ and X-ray emission) in the thick disks of
galaxies imply the presence of distinct interstellar phases.  Howk \&
Savage (2000) have compared high resolution (sub-arcsecond) imaging of
\halpha\ from the extraplanar DIG and of absorption due to the
extraplanar dust in NGC 891.  This comparison revealed not only that
the structures seen in dust and \halpha\ are not physically related,
but that the WIM is much more smoothly distributed than the material
traced by dust extinction.  This almost certainly implies the volume
filling factor of the material traced by extraplanar dust is much
smaller than that of the DIG.  Rossa and collaborators (this
proceedings; Rossa et al.  2004) have recently presented even
higher-resolution images of NGC 891 acquired with {\em HST} that lead
to the same conclusion: the dust and the ionized gas occupy separate
regions of space with very little correspondence and, therefore,
represent {\em distinct phases} of the multiphase thick disk in this
galaxy.

There is also some evidence for CO emission, a direct tracer of CNM
material, in the thick disks of spiral galaxies.  Garc\'{i}a-Burillo
et al. (1999) have presented CO observations of the edge-on galaxy NGC
4013.  Their interferometric maps reveal the presence of CO-bearing
extraplanar filaments, some of which are coincident with extraplanar
dust seen by Howk \& Savage (1999).  This is a direct indication of
the presence of a CNM in the thick disk of this galaxy.
Unfortunately, few galaxies have been observed with the sensitivity
and resolution required to detect these structures.  NGC 891 has been
mapped in CO, although the picture in this galaxy is far from clear:
single-dish (Garc\'{i}a-Burillo et al. 1992) and interferometric
(Scoville et al. 1993) observations give conflicting results.  We are
pursuing deep interferometric CO mapping of this galaxy to limit the
amount of CO in the thick disk of this system.

Perhaps the least direct, but most interesting, indicator for the
presence of a CNM in the thick disks of galaxies is the recent
evidence for thick disk star formation in spiral galaxies.  Several
authors have noted the presence of extraplanar \HII\ regions in the
thick disks of galaxies (Walterbos 1991; Ferguson, Wyse, \& Gallagher
1996; Howk \& Savage 1997, 2000).  These regions are too far from the
midplanes of the host galaxies for the OB associations required to
ionize the gas to have been born in the disk and subsequently ejected.
The study of the abundances and stellar content of such regions has
just begun (T\"{u}llman et al. 2003) and may give us important
information on the circulation of metals within the thick disks of
galaxies.  Indirectly, the presence of newly-formed hot stars in the
thick disks implies the presence of a CNM, since the latter is a
crucial ingredient for star formation.  We consider it likely that the
young stars have formed from clouds similar to those seen in our
images.

\section{Implications for the Nature of Extraplanar Matter}

\begin{figure}[!ht]
\plotone{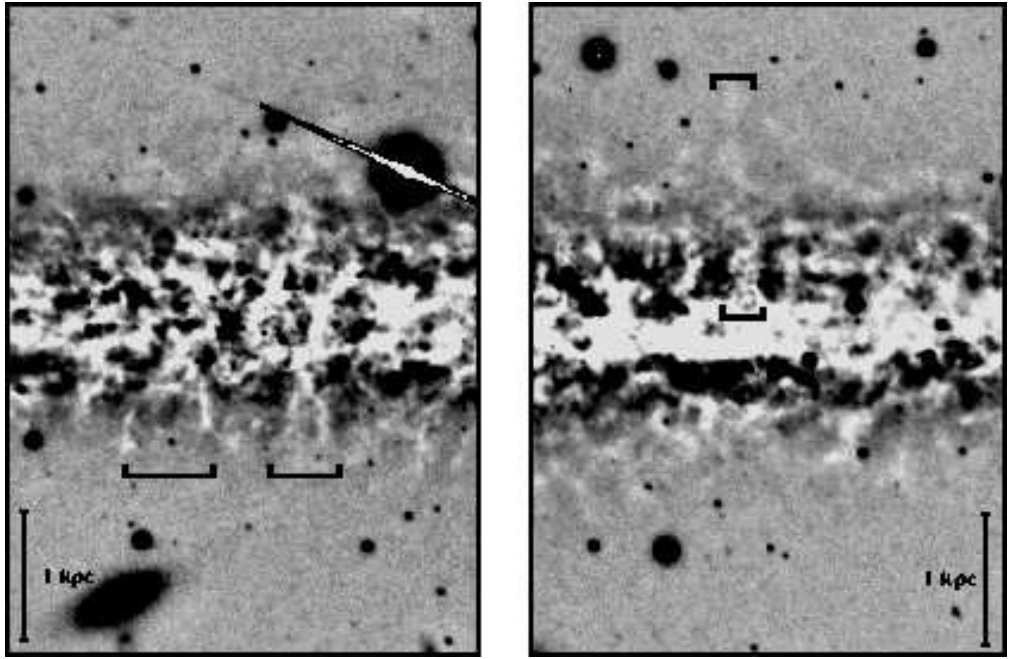}
\caption{Sections of the unsharp-masked V-band image of NGC 891 from Howk \&
  Savage (2000); these regions are centered to the NE (left) and SW
  (right) of the galaxy center.  The high-\z\ structures marked in
  these images are clearly connected to the disk of the galaxy.  The
  source of thick disk dust observed in this and other galaxies is
  expulsion from the thin disk.  The connection to the thin disk is
  not unique to NGC 891, but is also seen in other galaxies (e.g.,
  Thompson et al. 2004; Sofue et al.  1994). \label{fig:diskhalo}
}
\end{figure}

The presence of dust in the thick disks of galaxies has important
implications for understanding the nature of extraplanar gas in spiral
galaxies.  Perhaps the most fundamental implication of significant
amounts of dust in the thick disks of spiral galaxies is that much of
the interstellar material in the thick disk in these systems has been
expelled from the underlying thin disk rather than accreted from a
reservoir of primordial material.  This is strongly suggested by the
large amounts of dust directly visible in our images.  Furthermore,
while not generally true, there are indeed structures connecting the
extraplanar dust to the underlying thin disk (Figure
\ref{fig:diskhalo}).  The precise mechanism for expelling gas and dust
from the thin disk is not constrained, but some aspect of feedback
from massive stars likely drives the expulsion.

%%%%%%%%%%%%%%%%%%%%%%%%%%%%%%%%%%%%%%%%%%%%%%%%%%%%%%%%%%%%%%%%%%%%%%

%%%%%%%%%%%%%%%%%%%%%%%%%%%%%%%%%%%%%%%%%%%%%%%%%%%%%%%%%%%%%%%%%%%%%%

The statement that much of the thick disk ISM must have been expelled
from the thin disk does not necessarily constrain the nature of
``halo'' gas, material at very large distances from the plane.  The
maximum extent of dusty material observable through its optical
absorption against background starlight is $z\sim2$ kpc.  This varies
slightly among the observed galaxies and potentially with position in
an individual galaxy.  However, in the best studied case of NGC 891,
there is sufficient light from the stellar bulge and thick disk that
clumpy dust structures could have been seen to much larger heights in
the central regions (Howk \& Savage 2000).

The lack of detectable dust at larger heights has a few potential
causes.  Having argued the strongly-clumped dust at $z\la2$ kpc is
associated with a CNM, we have suggested that the interstellar
pressures at larger heights may be insufficient to support a CNM (Howk
\& Savage 1999, 2000).  In which case, the dust at these heights is
associated with a diffuse medium with little in the way of small
scale, high column density structures that would be detectable via
direct optical imaging.  Alternatively, there could simply be a lack
of dust at high-\z, either due to the smaller amount of gas at such
heights or a changing gas-to-dust ratio.

While the current observations do not allow us to determine which of
these scenarios is more likely, up-coming observations with the {\em
  Spitzer Space Telescope} and the {\em Galaxy Evolution Explorer} may
help by revealing a smooth component of dust at $z\ga2$ kpc in
galaxies.  The distinction is potentially important, as it could bear
on the amount of gas contributed to modern galaxy halos by on-going
infall of primordial material and on the possibility that dust (and
potentially gas) may escape a galaxy's potential altogether.

It is worth noting, also, that the presence of dust grains in the
thick disks of galaxies implies that the mechanisms that transport
material from the thin to thick disks are not sufficiently violent to
completely destroy the dust.  Thus, if feedback processes are
responsible for expelling the dust and gas into the thick disks of
galaxies, the characteristic velocities with which much of the present
thick disk matter is expelled was likely too small to produce much
dust destruction.  This suggests that much of the extraplanar material
may be simply displaced disk gas rather than material that started as
shock-heated gas and metal-enriched supernova ejecta.  The lifting of
material through magnetically-driven mechanisms (e.g,. the Parker
instability) or other more quiescent mechanisms may be important,
although they are not necessary.

\acknowledgements{Thanks to my collaborators T.W.J. Thompson (UCSD)
  and B.D. Savage (Wisconsin) for their contributions to this work.}

\end{document}